# Radiation hardness and post irradiation regeneration behavior of GaInAsP solar cells

R. Lang[1,*], J. Schön[1,2], J. Lefèvre[3], B. Boizot[3], F. Dimroth[1], D. Lackner[1]

1) Fraunhofer Institute for Solar Energy Systems, Heidenhofstr. 2, 79110 Freiburg, Germany

2) Laboratory for Photovoltaic Energy Conversion, University of Freiburg, Emmy-Noether-Str. 2, 79110 Freiburg, Germany

3) Laboratoire des Solides Irradiés, CEA – CNRS – Ecole Polytechnique, Institut Polytechnique de Paris, Palaiseau, France

*robin.lang@ise.fraunhofer.de

**Abstract**

Recent developments have renewed the demand for solar cells with increased tolerance to radiation damage. To investigate the specific irradiation damage of 1 MeV electron irradiation in GaInAsP lattice matched to InP for varying In and P contents, a simulation based analysis is employed: by fitting the quantum efficiency and open-circuit voltage simultaneously before and after irradiation, the induced changes in lifetime are detected. Furthermore, the reduction of irradiation damage during regeneration under typical satellite operating conditions for GEO missions (60°C and AM0 illumination) is investigated. A clear decrease of the radiation damage is observed after post irradiation regeneration. This regeneration effect is stronger for increasing InP-fraction. It is demonstrated that the irradiation induced defect recombination coefficient for irradiation with 1 MeV electrons after regeneration for 216 hours can be described with a linear function of InP-fraction between $1 \cdot 10^{-5}$ cm$^2$/s for GaAs and $7 \cdot 10^{-7}$ cm$^2$/s for InP. The results show that GaInAsP is a promising material for radiation hard space solar cells.

## 1. Introduction

Solar cells used in space applications are operated in an atmosphere of high energy particles, which will decrease the performance during the course of their operation. Therefore, solar cell materials with high radiation hardness are required. This demand will increase in the future, since recently electric orbit-raising has been successfully applied in modern telecommunication satellites to save launch cost. However, this technique comes with the downside of a significantly increased transit time through the Van Allen belt and thus increased irradiation damage to the solar cells even before reaching final orbit and payload operation.

The radiation hardness and regeneration properties of indium phosphide (InP) for low operating temperatures < 100 °C significantly exceed those of gallium arsenide (GaAs) and it was found, that this is also true for InP-based solar cell materials like GaInP and GaInAsP [1]. Therefore, it was proposed that the radiation hardness of such materials is related to its InP-fraction [2]. In this study, we investigate this assumption within the quaternary GaInAsP material system, where both the band gap and the InP-fraction can be varied in a wide range. Three GaInAsP solar cells with different composition and hence different InP-fraction and band gaps were grown on InP substrates to investigate the assumption. Since the radiation hardness not only depends on the initial defect creation due to the irradiation, but also on the subsequent defect regeneration ability of the specific semiconductor material, the defect regeneration also often referred to as annealing properties was studied in detail as a function of time and composition. The chosen regeneration conditions were based on typical operating conditions for communication satellites in geo-stationary orbits, namely 60°C and AM0 illumination.

In current space solar cells, it is common to use As-based alloys such as GaAs and GaInAs. Although these alloys can be grown with a very high quality, the resulting solar cells are typically the radiation weakest sub-cells in the multi-junction stack. Alloys from the quaternary GaInAsP system with its wide range of band gap and lattice constant combinations have the potential to replace the GaAs and GaInAs junctions with solar cells of the same band gap but with higher radiation hardness. In this work, the focus is on the general investigation of the material. The implementation of GaInAsP into multi-junction space solar cells is discussed elsewhere [3–5].

## 2. Approach

### 2.1 Experimental methods

The GaInAsP solar cells were grown by MOVPE with a multi-wafer AIX2800G4-TM reactor on 4" InP substrates using standard precursors and growth conditions [6]. For this study, three different $Ga_xIn_{1-x}As_yP_{1-y}$ solar cells with different band gaps of 0.9, 1.0, and 1.1 eV were grown lattice matched to InP. The growth rate and hence the thickness of the absorber layers was derived from the oscillations in the EpiTT *in-situ* reflection of a 633 nm diode. Zn and Si were used as dopands in the active pn-junctions. The dopant levels were determined via electrochemical capacitance-voltage (ECV) measurements, whereas the composition was identified through lattice constant and band gap, measured by high resolution x-ray diffraction (HR-XRD) and derived from EQE measurements [7], respectively. These band gaps refer to compositions of $Ga_{0.31}In_{0.69}As_{0.67}P_{0.33}$, $Ga_{0.23}In_{0.77}As_{0.49}P_{0.51}$, and $Ga_{0.16}In_{0.84}As_{0.34}P_{0.66}$. We did not see a significant influence on BOL solar cell performance due to the different compositions in our analysis of the material.

| Layer | Thickness, Doping |
|---|---|
| n-InP Cap | 50 nm, $-1\cdot10^{19}$ cm$^{-3}$ |
| n-InP Window | 350 nm, $-5\cdot10^{18}$ cm$^{-3}$ |
| n-In$_x$Ga$_{1-x}$As$_y$P$_{1-y}$ Emitter | 100 nm, $-1.5\cdot10^{18}$ cm$^{-3}$ |
| p-In$_x$Ga$_{1-x}$As$_y$P$_{1-y}$ Base | 1700 nm, $+9\cdot10^{16}$ cm$^{-3}$ |
| p-InP Nucleation | 500 nm, $+1\cdot10^{18}$ cm$^{-3}$ |
| p-InP Substrat | |

Fig. 1: Layer structure of the different GaInAsP solar cells. Emitter and base were grown with three different compositions.

The solar cell structure is displayed in Fig. 1. An absorption limited rather than a diffusion limited begin-of-life (BOL) design was chosen to be more sensitive to the bulk minority carrier lifetime and thus be more robust against uncertainties in the measurements and the input parameters. The high performance of the design is known from previous work [4]. According to our simulations, the relative effect of regeneration is almost the same as for an end-of-life (EOL) design. Due to the small and highly n-doped emitter, the investigation is mostly sensitive to the p-type material of the thick base layer.

The base thickness of the GaInAsP cells with band gaps of 0.9 and 1.0 eV is 1700 nm but due to a mistake in the growth time, the thickness of the 1.1 eV solar cells is only 1350 nm. The thinner absorber has no influence on the defect recombination coefficient explained in section 2.2. The epitaxy wafers were processed into 4 cm² solar cells by photolithography and wet chemical etching. No anti-reflection coating (ARC) was applied. The reference GaAs cell had an EOL design with a 1.7 µm thick absorber.

All IV-measurements were conducted in a WaveLabs LED array sun simulator with a spectrum close to the AM0 spectrum. In this setup, it is possible to vary the chuck temperature from 10 to 80 °C and automatically measure the IV-curve at defined intervals. The same sun simulator was used for the regeneration experiments at a solar cell temperature of 60 °C and under illumination with the AM0 spectrum, which is close to typical space operating conditions (SOC). In contrast to the actual operating conditions, the cell is not kept at the maximum power point during the regeneration time, but at $V_{OC}$. These regeneration conditions are slightly different from the ECSS standard [8], where the cell is first illuminated for 2 days under AM0 at 25 °C, followed by 1 day in the dark at 60 °C. The difference between both conditions will be investigated in this study (Fig. 5). EQE-measurements were conducted using the so-called differential spectral responsivity method [9], where frequency modulated monochromatic light and continuous bias light are used together as test light and the signal related to the monochromatic light is detected with a lock-in technique. The EQE measurements were used to calculate $J_{SC}$ at 25 °C.

The electron irradiations took place in the SIRIUS irradiation facility at the Laboratoire des Solides Irradiés (Ecole Polytechnique, France). The samples were irradiated at room temperature under a He-atmosphere on a water cooled sample holder with 1 MeV electrons at three different electron fluences of $3 \cdot 10^{14}$, $1 \cdot 10^{15}$, and $3 \cdot 10^{15}$ cm$^{-2}$. We chose 1 MeV electrons since they are typically used to rate the radiation hardness of a material, for example in the ECSS standard mentioned above.

## 2.2 Simulation method

Numerical device simulation of the solar cells allows for a material specific analysis of the irradiation damage. The optical and electrical modeling was performed using TCAD Sentaurus [10], which is a commercial software package that has previously been used to model GaAs [11, 12] and GaInP solar cells [13]. The transfer matrix method is utilized to simulate optical absorption in the different solar cell layers. The solar cells (see Fig. 1) are simulated with doping dependent mobilities [14], thermionic emission at heterointerfaces, and Fermi statistics within a symmetry element that consists of half a finger distance. Models and parameters for absorption, band structure, doping dependent mobility and intrinsic recombination in the GaAs cell is taken from Ref. [11, 12], whereas material parameters for GaInAsP are morphed between known materials. The charge carrier recombination is described by applying injection dependent models for radiative, Auger and Shockley-Read-Hall (SRH) recombination as well as surface recombination at the window interfaces. It is assumed that the electron irradiation primarily influences the SRH lifetime of electrons and holes while the irradiation effects on other parameters including the mobility are negligible for the cell performances. Thus the SRH lifetime of electrons and holes is the only unknown parameter in the models and can be determined by fitting measured quantum efficiency and $V_{OC}$ simultaneously. Both, quantum efficiency and $V_{OC}$ of the investigated solar cells are very sensitive to the non-radiative lifetime of the electrons in the base. The fitted low injection minority carrier lifetime in the base before and after irradiation are taken for further evaluation of the irradiation damage as described in the next section. The hole lifetime in the emitter is above 0.1 ns in all investigated solar cells and has only a minor influence on the $J_{SC}$.

## 2.3 Material specific irradiation damage

The remaining factor is often used to describe the radiation hardness of a material. It relates the EOL to the BOL performance by simply taking the ratio of both measured quantities. The remaining factor can be quite misleading when comparing different materials and solar cell designs. It depends strongly on the investigated cell structure and especially on the non-radiative recombination in the absorber materials before irradiation, i.e. the remaining factor increases with decreasing BOL material quality. In order to rate the effect of irradiation, we rather use the irradiation induced defect recombination coefficient $\kappa$ as defined below. The minority carrier lifetime before ($\tau_{BOL}$) and after irradiation ($\tau_{EOL}$) can be split in the lifetimes associated with radiative and non-radiative recombination:

$$\frac{1}{\tau_{BOL}} = \frac{1}{\tau_{rad}} + \frac{1}{\tau_{non-rad}(BOL)} \quad \text{and} \quad \frac{1}{\tau_{EOL}} = \frac{1}{\tau_{rad}} + \frac{1}{\tau_{non-rad}(EOL)} = \frac{1}{\tau_{rad}} + \frac{1}{\tau_{non-rad}(BOL)} + \frac{1}{\tau_{id}},$$

where $\tau_{id}$ denotes the minority carrier lifetime associated to defects introduced by the electron irradiation. This equation holds under the assumption that possible irradiation induced carrier removal is negligible for the investigated doses in comparison to initial doping of the layer as expected from [15]. Thus $\tau_{id}$ can be calculated by:

$$\tau_{id} = \left(1/\tau_{EOL} - 1/\tau_{BOL}\right)^{-1}. \tag{1}$$

The inverse of $\tau_{id}$ is proportional to the defect density created by the specific radiation dose for electrons at a specific kinetic energy. Thus by dividing through the electron fluence $\Phi_{e^-}$ we arrive at the irradiation induced defect recombination coefficient $\kappa$:

$$\kappa = 1/(\tau_{id} \cdot \Phi_{e^-}), \tag{2}$$

$\kappa$ is a material and doping polarity specific value, independent of the BOL material quality, the solar cell structure or irradiation dose. This quantity can be used to rate true "radiation hardness". The difference to the previously proposed damage coefficient [1] is that the damage coefficient is a function of the diffusion length and thus, dependent on the lifetime and the minority carrier mobility. The mobility differs widely for different materials, but changes very little with irradiation as for typical III-V semiconductors at room temperature the mobility is dominated by phonon scattering [16]. Thus we consider $\kappa$ to be more suitable to compare different materials or material systems.

## 3. Results

### 3.1 Degradation

Before electron irradiation, IV-curves and the IQE for the three different GaInAsP solar cells were measured (Fig. 2). As expected, a clear increase of the open-circuit voltage ($V_{OC}$) and a decrease of the short-circuit current density ($J_{SC}$) were measured with increasing band gap from 0.9 eV to 1.1 eV. The $V_{OC}$ and fill factor (FF) values as well as high internal quantum efficiencies above 95 % indicate high material qualities for all three cells. Part of the drop in $J_{SC}$ in the 1.1 eV cell is due to the thinner absorber compared to the other solar cells with lower band gap.

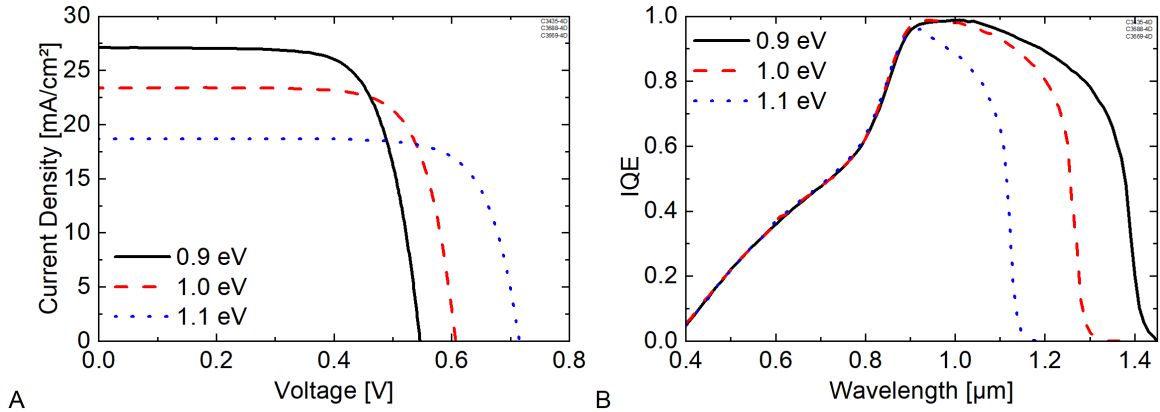

Fig. 2: BOL IV- (A) and IQE-curves (B) of 4 cm² GaInAsP solar cells with band gaps ranging from 0.9 to 1.1 eV. The IV-curves were measured under AM0 in a WaveLabs LED array sun simulator.

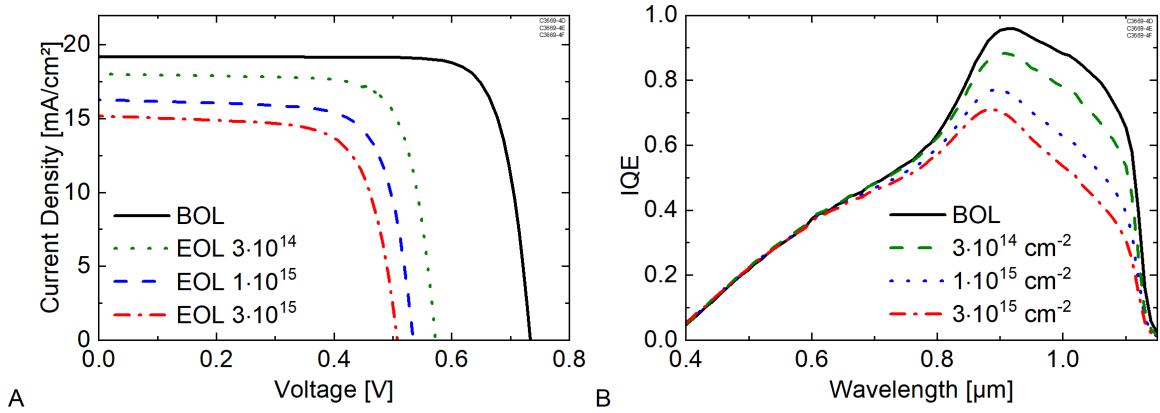

Fig. 3: BOL and EOL IV- (A) and IQE-curves (B) for the 1.1 eV GaInAsP solar cell at different fluences of $3 \cdot 10^{14}$, $1 \cdot 10^{15}$, and $3 \cdot 10^{15}$ cm$^{-2}$ of 1 MeV electrons. The IV-curves were measured under AM0 in a WaveLabs LED array sun simulator.

After irradiation with 1 MeV electrons at fluences of $3 \cdot 10^{14}$, $1 \cdot 10^{15}$ and $3 \cdot 10^{15}$ cm$^{-2}$ the same cells were measured again. Fig. 3 shows the measured IV and IQE curves after electron irradiation for the 1.1 eV solar cells as an example. A significant decrease of both $J_{SC}$ (IQE) and $V_{OC}$ was observed even for the lowest irradiation dose compared to the measurements before irradiation. The increase of the irradiation dose to $1 \cdot 10^{15}$ and $3 \cdot 10^{15}$ cm$^{-2}$ lead to further decrease of the solar cell characteristics. The remarkable drop of $V_{OC}$ already after the small irradiation dose of $3 \cdot 10^{14}$ is due to the high charge carrier lifetime of the GaInAsP before irradiation. $J_{SC}$ is less sensitive to lifetime changes as long as the charge carrier diffusion length is long enough to reach the pn-junction. Note that the InP window and contact operates like a spectral filter for wavelengths <0.92 μm. In Table 1 the $J_{SC}$ and $V_{OC}$ of all the solar cells are reported.

## 3.2 Thermal and Illumination induced Irradiation Defect Regeneration

The solar cells irradiated with an electron fluence of $\Phi_{e^-} = 1 \cdot 10^{15}$ cm$^{-2}$ were annealed under "space operating conditions" (60 °C and AM0) for several days.

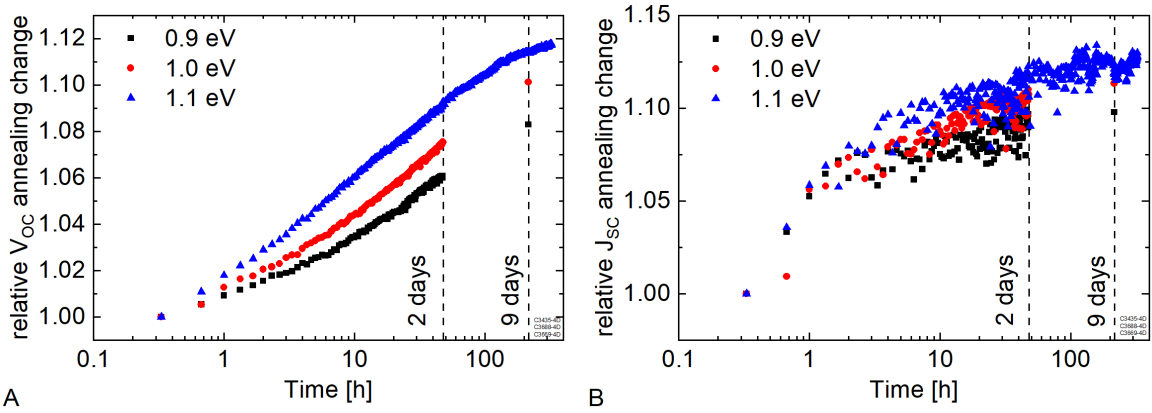

Fig. 4: Relative change of $V_{OC}$ (A) and $J_{SC}$ (B) during regeneration for 48 hours under AM0 at 60 °C for the 0.9 and 1.0 eV GaInAsP compositions and during more than 320 hours for the 1.1 eV composition. The starting point normalized to 1 represents the EOL values after a 1 MeV electron fluence of $1 \cdot 10^{15}$ cm$^{-2}$.

Fig. 4 shows the *in-situ* measurements of $V_{OC}$ and $J_{SC}$ during 2 days of regeneration under AM0 at 60 °C for the 0.9 and 1.0 eV GaInAsP compositions and during 2 weeks for the 1.1 eV composition. The results are plotted relative to the corresponding EOL values. IV measurements were taken every 20 minutes and the results are plotted with a logarithmic time scale. The increase in $V_{OC}$ for the first 2 days shows the exponential behavior of the regeneration (see Fig. 3A). As expected, a higher InP-fraction (higher band gap) leads to a significantly higher regeneration effect. After 2 days, the $V_{OC}$ recovery has still not saturated. The $V_{OC}$ of the 1.1 eV sample starts to saturate after 1 week but continues to increase further at a lower rate even after 2 weeks. As a result of this experiment, the 0.9 and 1.0 eV samples were also regenerated for an additional week; 9 days in total and showed a similar increase in voltage. The $J_{SC}$ of the samples behaves differently and increases mainly within the first hours of the experiment. Afterwards, it continues to increase but more slowly (see also Table 1).

Table 1: Overview of the BOL, EOL ($\Phi_{e^-} = 1 \cdot 10^{15}$ cm$^{-2}$), and regenerated (under AM0 at 60 °C for 2 and 9 days) solar cell key parameters as well as $\tau_{id}$ for a GaAs solar cell and the three different GaInAsP solar cells investigated in this study.

| Base Material | $E_g$ [eV] | State | $V_{OC}$ [mV] | $W_{OC}$ [mV] | $J_{SC}$ [mA/cm²] | η [%] | $\tau_{id}$ [ns] |
|---|---|---|---|---|---|---|---|
| Ga$_{0.31}$In$_{0.69}$As$_{0.67}$P$_{0.33}$ | 0.90 | BOL | 544 | 356 | 27.3 | 11.2 | |
| | | EOL $1 \cdot 10^{15}$ | 359 | 541 | 21.9 | 5.3 | 0.09 |
| | | 2d Reg. | 374 | 526 | 23.4 | 6.1 | |
| | | 9d Reg. | 382 | 518 | 24.3 | 6.4 | 0.2 |
| Ga$_{0.23}$In$_{0.77}$As$_{0.49}$P$_{0.51}$ | 0.99 | BOL | 617 | 373 | 24.3 | 11.7 | |
| | | EOL $1 \cdot 10^{15}$ | 432 | 558 | 20.1 | 6.3 | 0.075 |
| | | 2d Reg. | 462 | 528 | 22.2 | 7.5 | |
| | | 9d Reg. | 472 | 518 | 22.9 | 7.9 | 0.25 |
| Ga$_{0.16}$In$_{0.84}$As$_{0.34}$P$_{0.66}$ | 1.11 | BOL | 733 | 377 | 19.2 | 11.5 | |
| | | EOL $1 \cdot 10^{15}$ | 534 | 576 | 16.3 | 6.4 | 0.05 |
| | | 2d Reg. | 591 | 519 | 18.6 | 8.5 | |
| | | 9d Reg. | 600 | 510 | 18.7 | 8.7 | 0.3 |
| GaAs | 1.42 | BOL | 1054 | 366 | 34.6 | 30.8 | |
| | | EOL $1 \cdot 10^{15}$ | 922 | 498 | 32.8 | 25.1 | 0.1 |
| | | 2d Reg. | 922 | 498 | 32.5 | 24.9 | 0.1 |

In Table 1 the key parameters of the different GaInAsP solar cells and a GaAs solar cell after regeneration are compared to the BOL and EOL measurements. The $W_{OC}(= E_g/q - V_{OC})$, with the band gap energy $E_g$, is often used as a measure to rate the material quality of solar cells. Before irradiation, it is well below 400 mV for all three GaInAsP compositions and for the GaAs solar cell, an indication for high material quality solar cells within this band gap range [17]. The BOL $W_{OC}$ difference between the different GaInAsP band gaps is due to the dependence of the radiative limit on the band gap and not due to an actual difference in material quality. After the strong $W_{OC}$ increase of almost 200 mV caused by the irradiation defects, the $W_{OC}$ decreases again during regeneration for all three GaInAsP solar cells. In contrast, GaAs is not affected by the 2 days regeneration procedure as expected from literature [2]. Note that the high $J_{SC}$ and the high remaining factor of the GaAs cell is due to a double layer ARC and an optimized EOL cell design. As mentioned before, in contrast to the remaining factor, $\kappa$ and therefore also $\tau_{id}$ is purely a material property.

### 3.3 Comparison with ECSS Standard

The 2 days regeneration under AM0 at 60 °C differs from the regeneration standard set by the European Cooperation for Space Standardization (ECSS). The ECSS standard consists of keeping the solar cells illuminated with the AM0 spectrum for 2 days at 25 °C and afterwards 1 day in the dark at 60 °C, in our study we keep them under AM0 at 60 °C simultaneously [8].

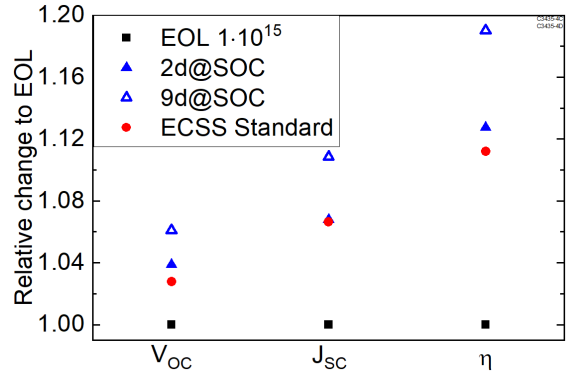

Fig. 5: Relative change of $V_{OC}$ and $J_{SC}$ after 1 MeV electron irradiation with a fluence of $1 \cdot 10^{15}$ cm$^{-2}$ for the 0.9 eV GaInAsP solar cells compared to the EOL values for different regeneration conditions: Under AM0 at 60 °C (after 2 and 9 days) and at the ECSS standard conditions.

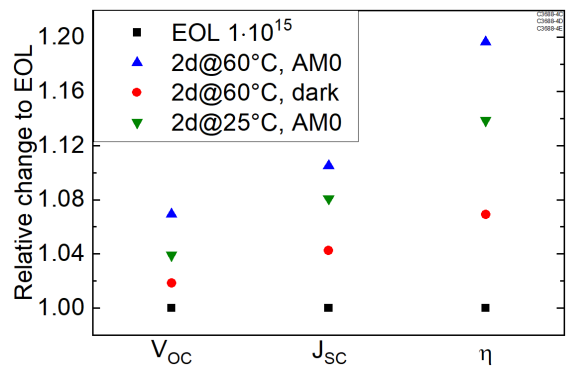

Fig. 6: Relative change of $V_{OC}$, $J_{SC}$, and $\eta$ after 1 MeV electron irradiation with a fluence of $1 \cdot 10^{15}$ cm$^{-2}$ for the 1.0 eV GaInAsP solar cells compared to the EOL values for different regeneration conditions: 2 days under AM0 at 60 °C, 2 days at 60 °C in the dark and 2 days under AM0 at 25 °C.

Both regeneration methods are compared in Fig. 5 along with 9 days of regeneration under AM0 at 60 °C. It is shown how $V_{OC}$ and $J_{SC}$ change with respect to their EOL ($\Phi_{e^-} = 1 \cdot 10^{15}$ cm$^{-2}$) values for the 0.9 eV GaInAsP solar cell. The $V_{OC}$ regeneration after 2 days under AM0 at 60 °C is slightly superior to the ECSS standard, whereas there is no measurable difference in $J_{SC}$. The $V_{OC}$ regeneration after 9 days under AM0 at 60 °C is about 3% higher than the value produced by the standard procedure. Note that this difference is already seen for the sample which shows the weakest regeneration. Therefore, in case of GaInAsP, the ECSS standard underestimates the solar cell regeneration capacity and consequently the expected solar generator performance during a space mission.

In order to differentiate the influence of temperature and illumination on the regeneration behavior, the 1.0 eV GaInAsP cells were annealed always for two days but under different conditions: under AM0 at 60 °C, at 60°C in the dark and under AM0 at 25 °C (Fig. 6). Each condition leads to a different measurable defect regeneration of the solar cells. The change in $V_{OC}$ after 2 days indicates that the light induced free carrier injection has a stronger impact on the regeneration process than the temperature. The effect is known for InP and GaInP [18], defects can capture electrons and the subsequent recombination significantly reduces the necessary regeneration energy. Compared to the overall heating of the sample, this is a specific, localized method of energy transfer directly at the defect. After 2 days, the combination of light and temperature has the largest effect. This also serves to explain the small difference between AM0 at 60 °C and ECSS. However, since the regeneration was not saturated after 14 days, it is unclear how the methods differ concerning the number of defects that can be cured.

### 3.4 Irradiation limited lifetime

Simulations were utilized to determine the SRH lifetime of GaInAsP for BOL, EOL ($\Phi_{e^-} = 1 \cdot 10^{15}$ cm$^{-2}$) and after regeneration by simultaneously fitting the measured IQE and $V_{OC}$ (see section 2.2 for details).

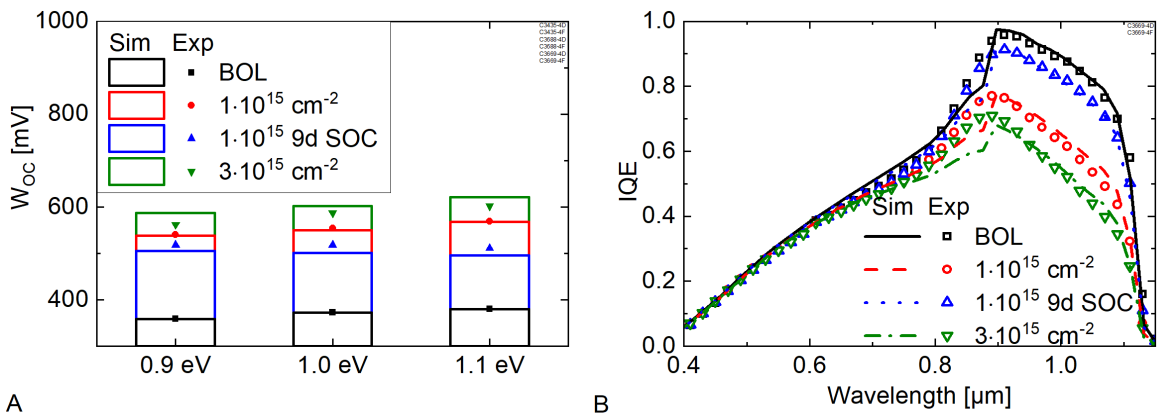

**Fig. 7:** Comparison of simulated and measured $W_{OC}$ of the different GaInAsP cells (A). The IQE is plotted exemplary for the 1.1 eV GaInAsP material in B. Results are shown for cells before and after irradiation with 1 MeV electron at a fluence of $1 \cdot 10^{15}$ and $3 \cdot 10^{15}$ cm$^{-2}$. For $\Phi_{e^-} = 1 \cdot 10^{15}$ cm$^{-2}$ the regenerated results after 9 days under AM0 at 60 °C is displayed as well (SOC). The same effective defect recombination coefficient $\kappa$ is assumed for different electron fluxes.

Fig. 7A shows the comparison of the simulated and the measured $W_{OC}$ values for the different GaInAsP solar cells. Fig. 7B shows the simulated and measured IQE exemplary for the 1.1 eV GaInAsP composition. The simulations for an electron fluence of $3 \cdot 10^{15}$ cm$^{-2}$ were conducted based on the $1 \cdot 10^{15}$ cm$^{-2}$ results by assuming a proportionality of $1/\tau_{id}$ and $\Phi_{e^-}$. Note that the results directly after an electron fluence of $1 \cdot 10^{15}$ cm$^{2}$ followed by 9 days of regeneration are close to the results after irradiation with an electron fluence of $3 \cdot 10^{14}$ cm$^{-2}$ (Fig. 3). Overall, the measurements and simulations are in good agreement. The determined $\tau_{id}$ values ($\Phi_{e^-} = 1 \cdot 10^{15}$ cm$^{-2}$) before and after nine days under AM0 at 60 °C regeneration are summarized in Table 1. In terms of $\tau_{id}$, GaAs is superior to GaInAsP after irradiation, but becomes inferior after regeneration.

## 4. Discussion of GaInAsP Radiation Hardness with Respect to InP-Fraction

To further analyze the irradiation damage of the different materials, the irradiation induced defect recombination coefficient is calculated according to equation (2). Fig. 8A shows an overview of the defect recombination coefficient $\kappa$ for GaInAsP with varying InP-fraction. In this study, the InP-fraction is defined as the ratio of InP within the GaInAsP crystal and therefore equivalent to the percentage of either In or P, whichever is lower. From 0.9 to 1.1 eV this InP-fraction of the GaInAsP solar cell increases from 33 to 66 % (see Table 1). Note that a low $\kappa$ denotes little damage due to electron irradiation. Before regeneration (filled red circles), we can see a clear trend towards higher $\kappa$ values with increasing InP-fraction. This is probably due to the increased In-fraction, since In has a higher atomic number $Z$ than Ga and As and the cross section of the high energy electron and matter interaction depends on $Z^2/M$, with the atomic mass $M$ [19]. Therefore, we expect to have an increased number of initial defects (mostly In-vacancies $V_{In}$ and In-interstitials $In_i$) in samples which contain more InP. The higher electron – In-atom collision probability and thus the higher defect density will decrease the SRH lifetime of the samples [20], which would explain the observed effect. However, the position of the defect levels of the created $V_{In}$ and $In_i$ and their electron capture cross section also influence the minority carrier lifetime. Therefore, not only the number of defects determines the radiation hardness. It is likely that the defect energy levels and capture cross-sections change with the InP-fraction. Without exact knowledge of the capture cross-section one could only speculate whether the increasing In-fraction and thus increased interaction with the electron irradiation or a change in the properties of the defects leads to the increased irradiation induced recombination rates with increasing InP-fraction in GaInAsP.

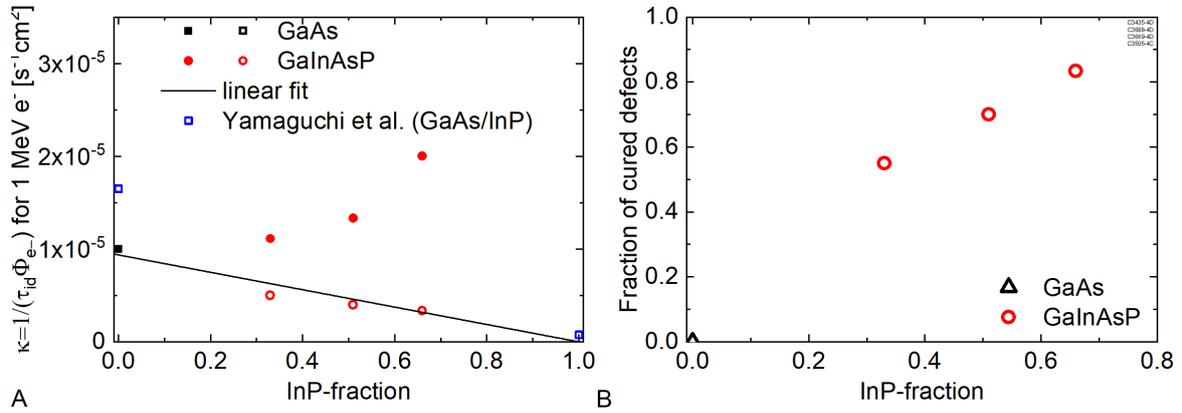

**Fig. 8: A:** Overview of the material specific recombination coefficient $\kappa$ for different p-type GaInAsP materials plotted over their respective InP-fraction both before (filled) and after (open) regeneration. The black line represents a linear fit through our experimental values with the constraint $\kappa \geq 0$. The blue open squares are calculated from Fig. 2 in [1] with the mobility values used for simulation of GaAs and InP in this work. **B:** The fraction of cured defects in the GaInAsP solar cells after irradiation with an electron fluence of $10^{15}$ cm$^{-2}$ and 9 days regeneration under AM0 at 60 °C as function of their InP-fraction.

However, after 9 days under AM0 at 60 °C (open red circles), $\kappa$ drops significantly for GaInAsP and the trend reverses: higher InP-fractions lead to lower $\kappa$, i.e. lower recombination rates due to irradiation damage. Thus, a higher InP-fraction is beneficial in terms of irradiation hardness, as was originally assumed and reported in literature [2]. The stronger regeneration effect for the higher band gap GaInAsP is already indicated by the stronger $V_{OC}$ increase under AM0 at 60 °C (see Table 1). The analysis based on $\kappa$ allows for quantification of this effect. Under the assumption that the lifetime before and after regeneration is dominated by the same defect, the fraction of cured irradiation defects can be calculated with $1/\tau_{id}$ after regeneration divided by $1/\tau_{id}$ before regeneration. In GaInAsP with a band gap of 0.9 eV more than 50 % of the irradiation damage ($\Phi_{e^-} = 1 \cdot 10^{15}$ cm$^{-2}$) is cured during 9 days of AM0 at 60 °C (see Fig. 7B) despite the relative small increase in V$_{OC}$ (+22 mV). For GaInAsP with higher band gaps, the regeneration is stronger and reaches more than 80 % for GaInAsP with 1.1 eV.

Electron irradiation mainly causes defects by collisions with atoms of the crystal lattice and their following displacement. This atom displacement creates vacancy-interstitial pairs [19]. Most of these pairs are separated by no more than a few interatomic distances and thus can recombine during regeneration [21]. The effect of the regeneration depends on the migration energy of the defects as well as the amount of energy provided through temperature, illumination, or other means. In general, the higher radiation resistance of InP-related materials compared to GaAs is explained by the difference in defect migration energies. For example, the migration energies of In- (V$_{In}$) and P-vacancies (V$_P$) in InP are 0.26 eV and 1.20 eV, respectively, whereas the Ga (V$_{Ga}$) and As-vacancy (V$_{As}$) migration energies in GaAs are much higher (1.79 eV for V$_{Ga}$ and 1.48 eV for V$_{As}$) [22].

Therefore, in contrast to GaInAsP, GaAs is not affected by AM0 at 60 °C regeneration. This finding is in agreement with Yamaguchi et al. who found several orders of magnitude higher regeneration rates for InP compared to GaAs at moderate temperatures [2]. The black line in Fig. 8A corresponds to a linear fit using our results for GaAs and GaInAsP with the constraint $\kappa \geq 0$. In Ref. [1] the same damage coefficients were found for GaInAs and GaAs which makes a linear fit for $\kappa$ reasonable. Literature results for GaAs and InP are also shown for comparison (blue squares). It should be noted that the GaAs literature value of $\kappa$ is nearly twice as high as our result [1].

## 5. Conclusion

In this paper, we have systematically analyzed the material degradation of GaInAsP solar cells under electron irradiation and the following regeneration at 60 °C and AM0 illumination. Numerical simulations of the solar cells were utilized to quantify the irradiation damage of 1 MeV electron irradiation and to determine the irradiation induced defect recombination coefficient ($\kappa$), a material specific parameter. It is shown that $\kappa$, i.e. the effective defect density, depends strongly on the composition of GaInAsP. Directly after electron irradiation $\kappa$ increases with increasing InP-fraction and is even higher than in a GaAs reference. However, the regeneration of the irradiation defects improves with increasing InP-fraction at temperatures and illumination intensities typical for GEO missions (60 °C operation temperature and AM0 illumination). For GaInAsP with an InP content of 66%, after 9 days under AM0 at 60 °C more than 80% of the irradiation induced defects are cured. As a consequence, GaInAsP becomes an "irradiation harder" material for increasing InP content after a few days of illumination at 60 °C. The initial hypothesis that $\kappa$ decreases linearly with InP-fraction from 9.7 10$^{-6}$ s$^{-1}$cm$^2$ (GaAs) to 7.5 10$^{-7}$ s$^{-1}$cm$^2$ (InP) for 1 MeV electron irradiation was confirmed for GaInAsP but only after regeneration. Further, the high absorption coefficients of GaInAsP reduce the thickness requirements on the absorber and increase the possible remaining factors for InP rich material even more. Thus GaInAsP, especially with high InP-fractions, is a promising material for the integration in future radiation hard space solar cells.

## Acknowledgements

The authors would like to thank the technology group for the processing of the solar cells, E. Schäffer, F. Martin, and the whole CalLab team for the solar cell measurements, and S. Stättner for operation assistance of the MOVPE processes.

The authors acknowledge funding by the European Union's Horizon 2020 Research and Innovation Programme, under Grant Agreement no EU/776362 (RadHard).